\documentclass[aps,pre,reprint,showpacs,superscriptaddress]{revtex4-1}
\usepackage{epsfig}

\begin{document}


\title{Gravity-driven clustering of inertial particles in turbulence}


\author{Yongnam Park}
\affiliation{Department of Mechanical Engineering, Yonsei University, 50 Yonsei-ro,
Seodaemun-gu, Seoul, Korea}
\author{Changhoon Lee}
\affiliation{Department of Computational Science and Engineering \& Department
of Mechanical Engineering,
Yonsei University, 50 Yonsei-ro, Seodamun-gu, Seoul, Korea}
\email[]{clee@yonsei.ac.kr}


\date{\today}

\begin{abstract}
We report a new kind of particle clustering caused purely by gravity, discovered in our simulation of particle-laden turbulence. Clustering in a vertical strip pattern forms when strong gravity acts on heavy particles. This phenomenon is explained by the skewness of the flow velocity gradient in the gravitational direction experienced by particles, which causes horizontal convergence of particles.
\end{abstract}


\maketitle

Clustering of inertial particles in fluid, also known as preferential concentration, is a well-known phenomenon that is often found in particle-laden turbulence when the particle response time is of the same order as the Kolmogorov time scale of turbulence\cite{Maxey, Squires, Reade, Balachandar, Monchaux, Grabowski}. The main mechanism of this phenomenon is the expulsion of inertial particles from rotating structures by centrifugal force  when the particles interact with the coherent rotational motion of fluids. Therefore, most particles are found in a region where rotational motion is weak but where straining motion is observed. However, clustering of particles is not found when particles are so light that they behave like fluid particles or so heavy that their interactions with the fluid are weak. We report a different kind of particle clustering that occurs when heavy particles settle due to gravity, observed in our direct numerical simulations of homogeneous isotropic turbulence laden with particles. This clustering has totally different characteristics from clustering in the absence of gravity. In this paper, we provide quantitative evidence supporting particle clustering and a plausible interpretation of its mechanism.

We performed direct numerical simulation of particle-laden turbulence by solving the Navier-Stokes equations with continuity,
\begin{eqnarray}
 \frac{\partial  u_i}{\partial  t}+ u_j \frac{ \partial  u_i}{ \partial  x_j} & = &
-\frac{1}{\rho_f}\frac{\partial p}{\partial x_i}+
\nu_f \frac{ \partial^2 u_i}{\partial x_j \partial x_j}+
f_i  ,   \label{eq:N-SE}   \\  
\frac{\partial u_i}{\partial x_i} & = & 0 , \label{eq:continuity}
\end{eqnarray}
where $u_i$ is the velocity of the fluid phase, $p$ is the pressure, $\rho_f$ is the density of fluid, $\nu_f$ is the viscosity of fluid, and $f_i$ is a random force required to maintain stationary turbulence. The diameters of the laden particles are assumed to be much smaller than the smallest flow length scale, the Kolmogorov length scale, such that a point-particle approach is adopted. The governing equations for particle motion are then
\begin{eqnarray}
 \frac{d x^p_i}{d t} & = & v_i \\
 \frac{d v_i}{d t} & = & \frac{u_i-v_i} {\tau_p}- g \delta_{i2},
\end{eqnarray}
with $x^p_i, v_i, u_i$ and $g$ denoting the particle position, particle velocity, fluid velocity at the particle location, and gravitational acceleration, respectively. $\tau_p (=d^2\rho_p/18\rho_f\nu_f)$ is the particle relaxation time, with $d$ and $\rho_p $ denoting the diameter  and the density of a particle, respectively. The effective nondimensional parameters are the Stokes number St$(=\tau_p/\tau_{\eta})$ and the normalized settling velocity W$(=g\tau_p/v_{\eta})$, or interchangeably, the Froude number Fr($=v_{\eta}/(g \tau_{\eta}) =$ St/W), with $\tau_{\eta}$ and $v_{\eta}$ denoting the Kolmogorov time and velocity scales, respectively. In the point particle approach, collisions between particles and the two-way interaction between particles and fluid are neglected.

Equations for fluid motion (Eqs. \ref{eq:N-SE}, \ref{eq:continuity}) are solved in a periodic cube $[0,2\pi]^3$ using a spectral method on $128^3$ grids and the flow Reynolds number Re$_{\lambda}=70$\cite{Jung, Choi}. The time marching scheme for fluid and particle motions is the third-order Runge-Kutta method\cite{Abou1, Abou2}, and the interpolation of fluid velocity at particle locations is carried out using the fourth-order Hermite interpolation\cite{Lee2004,ChoiJ}.

Figure 1 shows a snapshot of particle distribution in an $x-y$ plane, with the $-y$ direction denoting the gravitational direction for various values of the Stokes number and gravity number. Here, red-colored (gray-shaded) spots are the region where the enstrophy of the fluid is high, and thus the fluid has a strong rotation. When there is no gravity and St=1, a typical preferential concentration of the particles is observed, as shown in Fig. 1(a). Most of the clustering of particles is found in the region with no strong fluid rotation. As gravity is applied to particles of St=1 (Fig. 1(b)), the clustering effect seems to weaken slightly\cite{Jin}, and the particles tend to align with the gravitational direction. Due to intense settling, some particles pass through the intense rotation region, and the interaction between the settling particles and the rotational structure described in Ref. \cite{Davila} becomes weaker. When heavy particles with St=4 settle under strong gravity (W=20 or Fr=0.2), a vertical pattern of particle clustering is clearly observed, as shown in Fig. 1(d). The particles show a strong alignment in the gravitational direction, forming a strip pattern. When St=4, particle inertia is so large that particles are hardly influenced by the fluid motion due to the long response time; it is remarkable that heavy particles settling due to strong gravity display such an intense clustering phenomenon. Different pattern of clustering due to gravity is clearly discernible from the case without gravity (Fig. 1(c)). A similar kind of clustering under even stronger gravity was attributed to the periodicity of the computational domain in Ref. \cite{Woittiez}. To confirm that this phenomenon is not due to a numerical artifact such as the periodicity of the computational domain or the large-scale forcing for stationarity, simulations were carried out in a vertically longer domain and in decaying turbulence. Similar particle clustering was observed. It can be concluded that the clustering is caused purely by gravity because we did not consider particle collision or two-way interactions between the particles and fluid.
 \begin{figure}
 \includegraphics[width=0.5\textwidth]{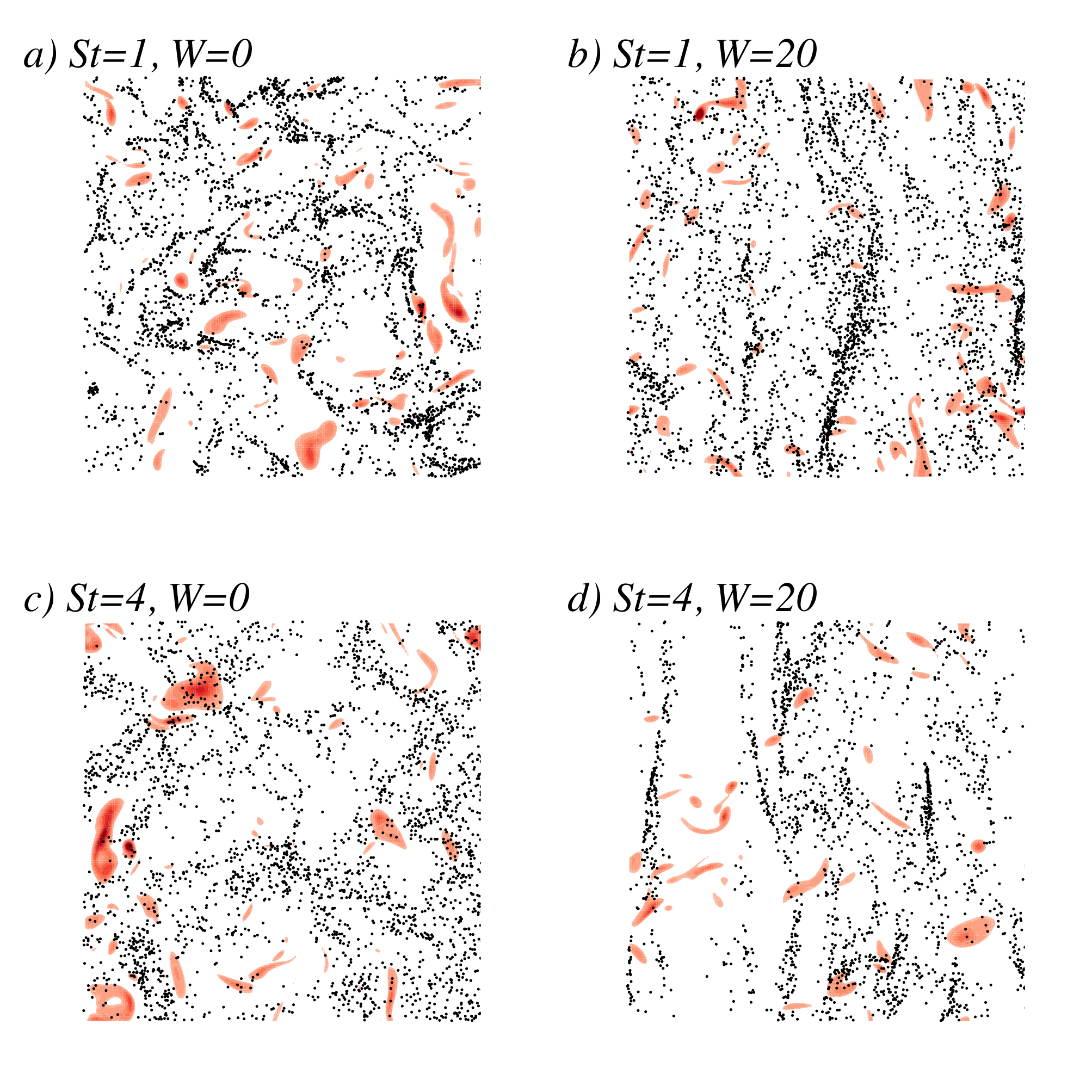}%
 \caption{(Color online) Particle distribution for various Stokes numbers and gravity numbers. Gravity acts in the vertical direction. Red (gray) regions denote intense enstrophy.  The corresponding Froude numbers are (a) Fr=$\infty$, (b) Fr=0.05, (c) Fr=$\infty$, (d) Fr=0.2.\label{fig1}}
 \end{figure}

In order to quantify the clustering phenomenon, we investigated the statistics of distances between particles. By analyzing the sorted distances from one particle to other particles, the distance to the closest particle, $\Delta$, and its horizontal and vertical components, $\Delta_x$ and $\Delta_y$, are easily identified. Although several methods to quantify clustering have been proposed\cite{Monchaux}, we selected the statistics of distance between particles to separately identify clustering in different directions. Figure 2 shows the averaged distances to the closest particle in each direction, $\overline{\Delta_x}$ and $\overline{\Delta_y}$,  normalized by the corresponding values for random distribution of the same number of particles for various values of St and W. Values smaller than 1 imply clustering, and deviation from the 45-degree line indicates anisotropic clustering. When gravity is weak, the clustering of particles for all Stokes numbers shows isotropic clustering. As gravity increases, however, $\overline{\Delta_y}$ becomes greater than $\overline{\Delta_x}$ for Stokes numbers greater than 1, clearly indicating that clustering tends to align with the gravitational direction. Gravity hardly influences particles with St=0.1 in terms of clustering. The clustering of particles with St=1, which occurs in the zero-gravity case, becomes weaker as gravity increases, although clustering tends to be anisotropic. Therefore, a new kind of vertical clustering, shown in Figs. 1(b) and 1(d), is observed only when St $\ge 1$.
 \begin{figure}
 \includegraphics[width=0.5\textwidth]{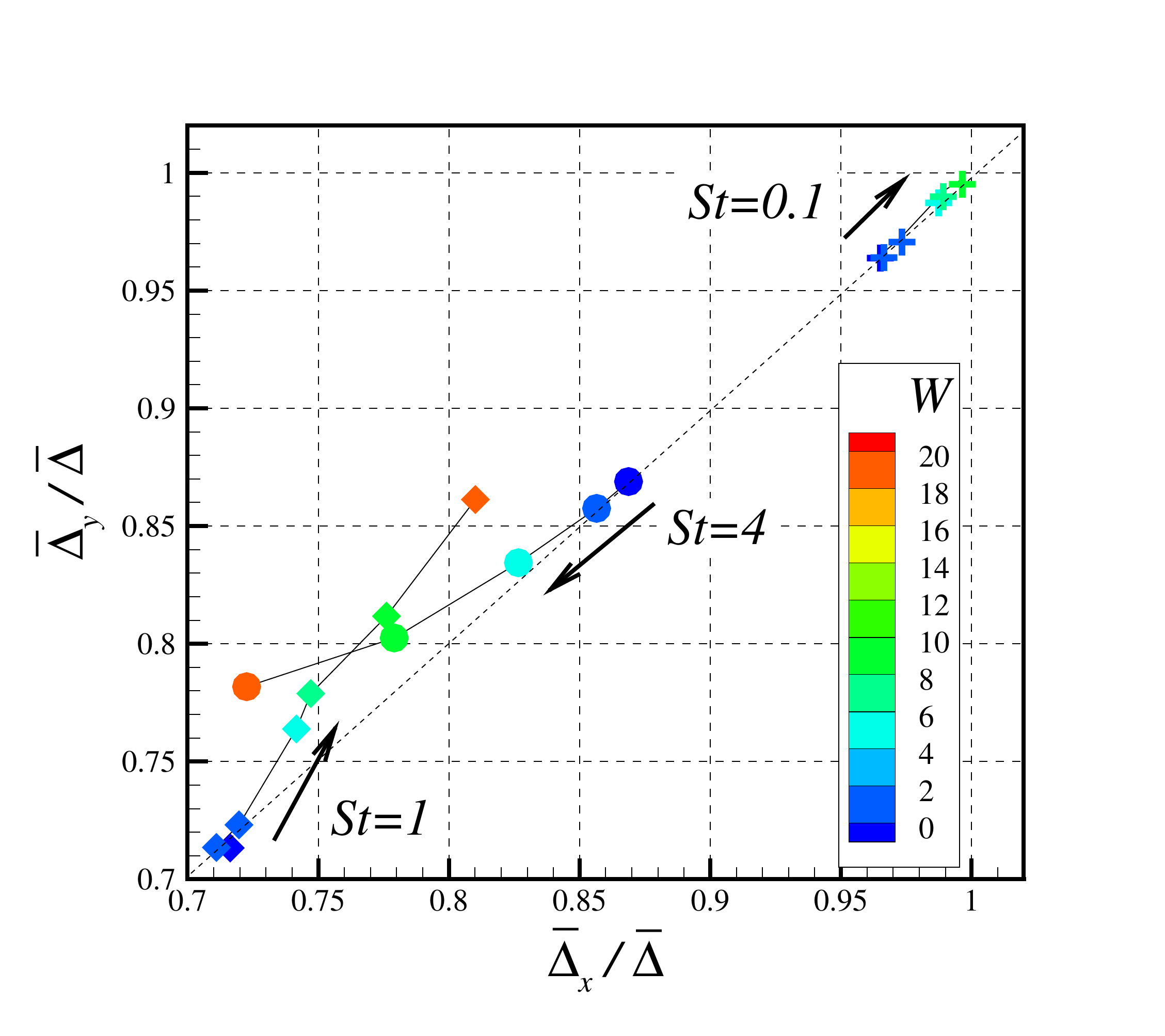}%
 \caption{(Color online) Average horizontal distance to the closest particle vs. average vertical distance for St=0.1 (cross), St=1 (solid square) and St=4 (solid circle). Color of symbols denotes gravity strength W as shown in the color legend. Each distance is normalized by the corresponding value for a random distribution of the same number of particles in the same domain so that values smaller than 1 indicate clustering.\label{fig2}}
 \end{figure}

To provide a plausible explanation for this kind of anisotropic clustering due to gravity, we focused on flow quantity, $\partial v/\partial y$, which is $-(\partial u/\partial x + \partial w/\partial z)$, the horizontal convergence of an incompressible fluid. It is well known that the skewness, $\overline{(\partial v/\partial y)^3}/\overline{(\partial v/\partial y)^2}^{3/2} $, which is $\sim -0.5$, for a wide range of Reynolds numbers in isotropic turbulence, is negatively skewed\cite{Belin}. This implies that a negative value of $\partial v/\partial y$ is more likely to have a larger magnitude than a positive value. The other interpretation of the negative skewness is that a negative value of $\partial v/\partial y$ is less frequently found in space than a positive value because integration of $\partial v/\partial y$ over the whole domain vanishes in homogeneous turbulence. Although the negative skewness of $\partial v/\partial y$ in turbulence is due to vortex stretching\cite{Batchelor47, Batchelor53, Frisch}, similar but more extreme asymmetry in $\partial v/\partial y$ can easily be found in the solution of the one-dimensional Burgers equation when a shock forms. $\partial v/\partial y$ of the flow field experienced by particles for St=1, 4 and W=0, 20 is shown in Fig. 3. The same kind of skewness of $\partial v/\partial y$ experienced by a particle can be observed as particles travel in space. The negative peaks of $\partial v/\partial y$ are more intense than the positive peaks, whereas the duration of the negative $\partial v/\partial y$ is shorter than that of the positive $\partial v/\partial y$. When gravity is effective, this asymmetry is preserved, and the durations of negative or positive $\partial v/\partial y$ become shorter due to the fast settling motion of particles, as shown in Fig. 3(b). When gravity is absent, this asymmetry will not affect the dispersion of particles because particles wander almost randomly in space. However, when gravity is present, particles settle vertically, and those exposed to more frequent positive convergence of the flow field will tend to cluster due to the Stokes drag. The effect of this asymmetry will be more pronounced as the particle Stokes number increases because particles with a longer relaxation time will respond more progressively to a longer duration of positive $\partial v/\partial y$. For example, rapid oscillatory variations of $\partial v/\partial y$ shown in the later part of Fig. 3(b) hardly affect the motion of heavy particles with St=4 (St=4 implies the response time of the particle $\tau_p = 4 \tau_{\eta}$). Recent observations of enhanced clustering of particles by gravity were attributed to either multiplicative amplification of many independent accelerations\cite{Gustavsson} or the preferential sweeping motion of particles\cite{Bec}.
 \begin{figure}
 \includegraphics[width=0.5\textwidth]{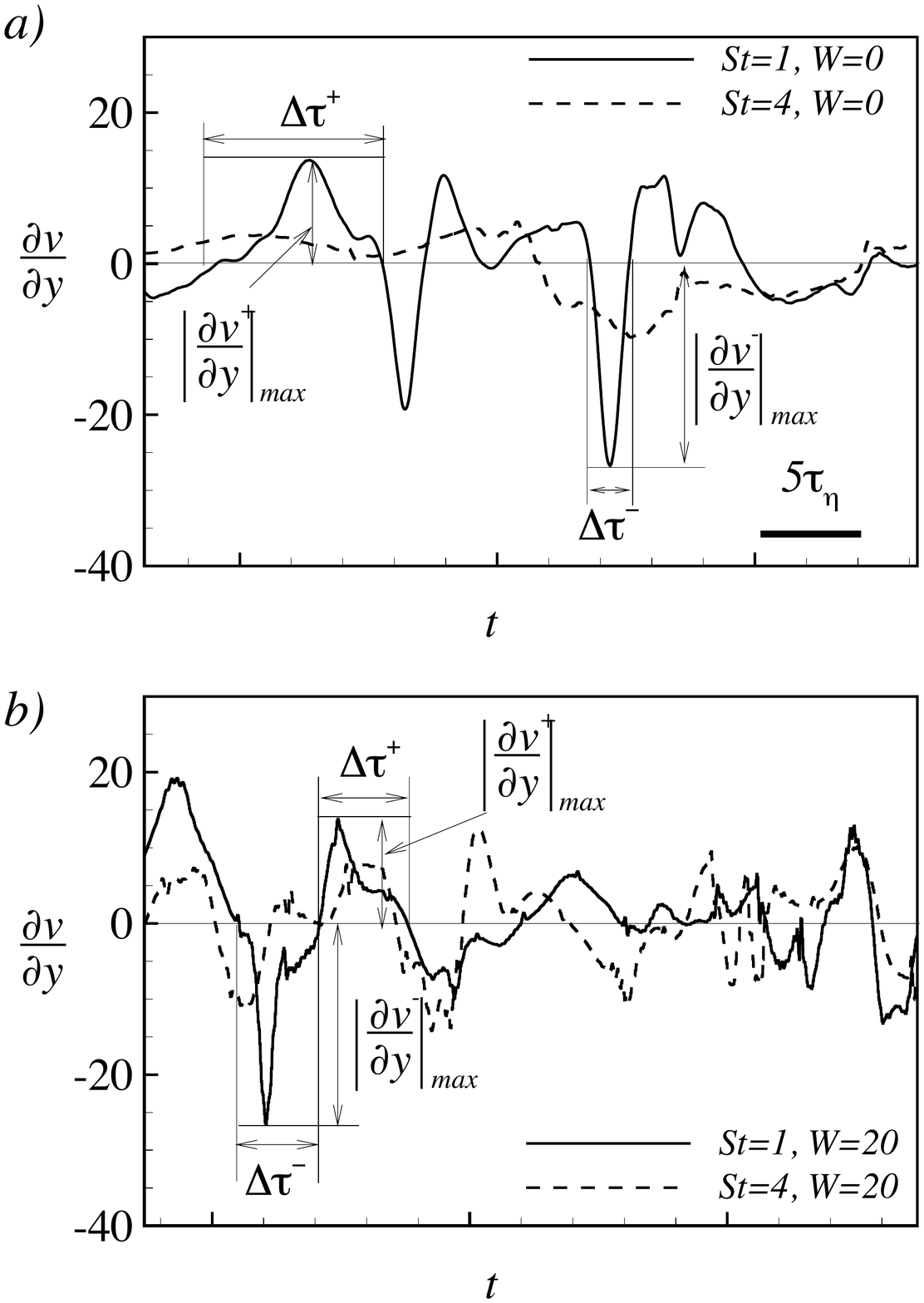}%
 \caption{$\frac{\partial v}{\partial y}$ of fluid experienced by a particle for St=1 and 4 (a) in the absence of gravity and (b) in the presence of gravity with W=20. $\Delta \tau^+$ and $\Delta \tau^-$ are the durations during which $\frac{\partial v}{\partial y}$ remains positive and negative, respectively. $|\frac{\partial v}{\partial y}|^+_{max}$ and $|\frac{\partial v}{\partial y}|^-_{max}$ are the peak values of $\frac{\partial v}{\partial y}$ during each period. The scale for $5\tau_{\eta}$ is given for comparison, with $\tau_p(=$St $\times \tau_{\eta})$. \label{fig3}}
 \end{figure}

Probability density functions(pdf) of the persistence time, $\Delta \tau^+$ and $\Delta \tau^-$ defined in Fig. 3, are shown in Fig. 4(a), clearly indicating a strong skewness. For all Stokes and gravity numbers, the persistence times are non-Gaussian and positively skewed. The tail part can be very well approximated by an exponentially decaying function, $P(\Delta \tau/\Delta \tau_{rms}) \simeq \alpha \exp (-\beta |\Delta \tau|/ \Delta \tau_{rms})$. As listed in Table 1, the estimated values of $\beta$ for $\Delta \tau^+$ and $\Delta \tau^-$ are very robust and clearly show a strong positive skewness. Compared to the no-gravity cases, the pdf in the presence of strong gravity is narrower due to the fast settling motion of the particles (the pdf in Fig. 4(a) is normalized by its own rms value). This increases the effectiveness of a longer persistence time in affecting the particles with a larger response time. A similar approach using the persistence time of the Okubo-Weiss parameter in two-dimensional turbulence showed that the PDF of the persistence time has exponential or power-law tails for the Eulerian or Lagrangian observations\cite{Perlekar}.

 \begin{figure}
 \includegraphics[width=0.5\textwidth]{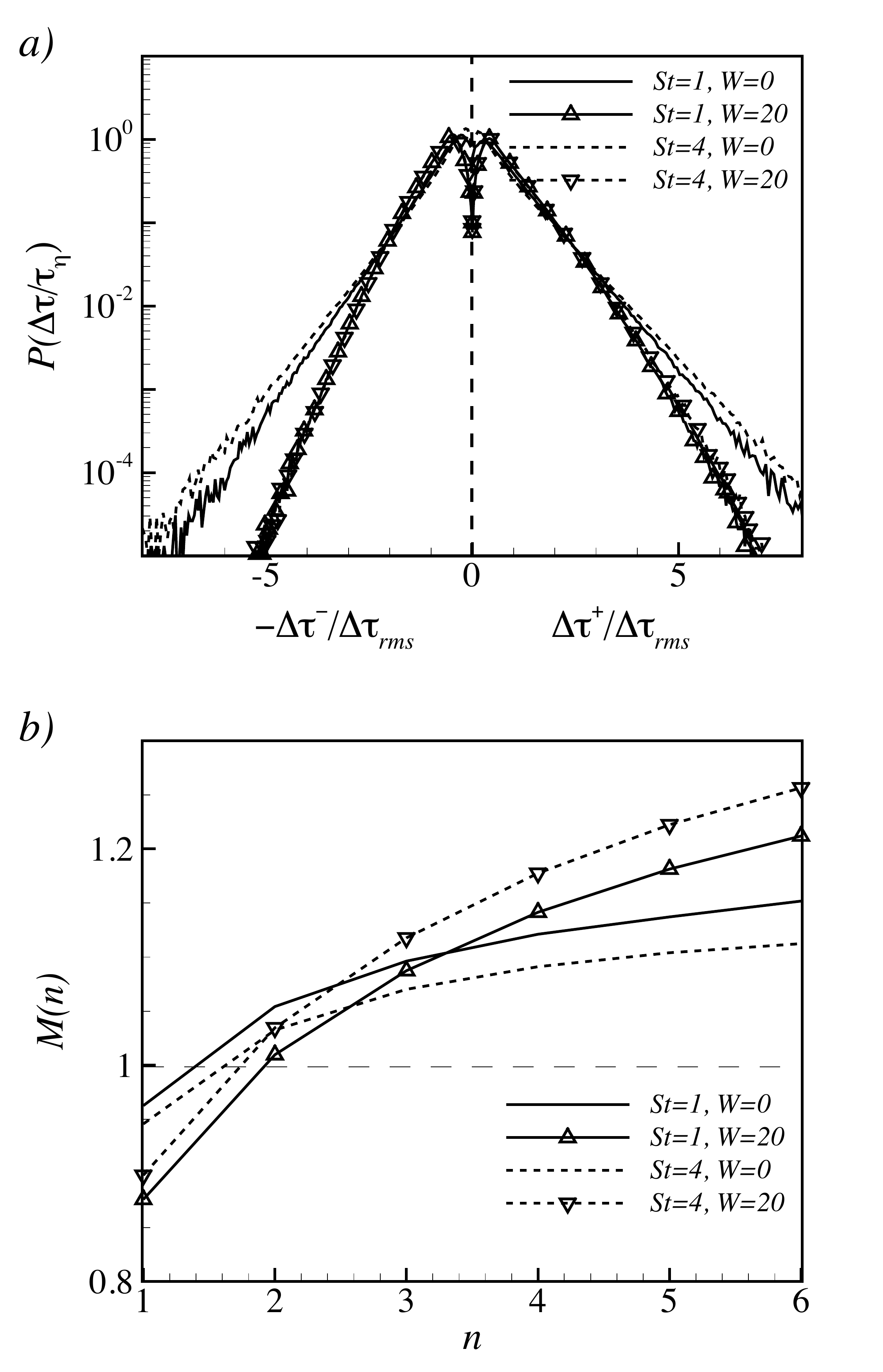}%
 \caption{(a) Probability density function of the persistence times, $\Delta \tau^+$ and $\Delta \tau^-$, normalized by its own $\Delta \tau_{rms}$ for St=1 and 4, and W=0 and 20. (b) The ratio of weighted $n$-th order moments of the persistence times defined in Eq. (\ref{eq:mn}) for the same parameters as in (a).\label{fig4}}
 \end{figure}
To quantitatively assess skewness between $\Delta \tau^+$ and $\Delta \tau^-$, we list the values of the normalized $n$-th order moment, ${\overline{|\Delta \tau/\Delta \tau_{rms}|^n}}^{1/n}$, obtained separately for $\Delta \tau^+$ and $\Delta \tau^-$ for $n=1 \sim 6$ in Table 1, clearly showing a stronger asymmetry as $n$ increases. Particularly, the ratio of weighted moments defined by
\begin{equation}
 M(n) = \left( \frac{\overline{ \left|\frac{\partial v}{\partial y}\right|^+_{max} (\Delta \tau^+)^n }}
                    {\overline{ \left|\frac{\partial v}{\partial y}\right|^-_{max} (\Delta \tau^-)^n }} \right)^{1/n} 
  \label{eq:mn}
\end{equation}
is shown in Fig. 4(b),
where $|\partial v/\partial y|^+_{max}$ and $|\partial v/\partial y|^-_{max}$ are the peak values of $|\partial v/\partial y|$ during each persistent time defined in Fig. 3. The reason for weighting the moments by the local peak value of $\partial v/\partial y$ is that the local value of $\partial v/\partial y$ and the duration affect the motion of the particles. Because $\int^T \partial v/\partial y~dt \simeq 0$ for a very long time $T$ in homogeneous turbulence, $M(1)$ is expected to be $\sim 1$. As shown in Fig. 4(b), $M(1)$ is $ \sim 1$ indeed. The ratio of moment $M(n)$ increases with $n$ for all cases, indicating that the skewness becomes more pronounced between longer persistence times, $\Delta \tau^+$ and $\Delta \tau^-$. This asymmetry, when strong gravity results in near vertical particle settling, causes a vertical strip pattern of clustered particles by exposing particles to horizontally converging fluid motion over a longer time. As the particle inertia increases, this asymmetry of longer persistence times influences particles more effectively due to a longer response time, and thus, vertical clustering is more recognizable. When gravity is absent, however, this asymmetry does not induce such a clustering because particles move in a random fashion.

\begin{table*}
\caption{\label{table1} Parameters $\alpha$ and $\beta$ of equation $P(\Delta \tau/\Delta \tau_{rms}) \sim \alpha \exp (-\beta |\Delta \tau |/\Delta \tau_{rms})$ fitting the pdf shown in Fig. 4(a), and normalized $n-$th order moments, ${\overline{|\Delta \tau/\Delta \tau_{rms}|^n}}^{1/n}$, for $n = 1 \sim 6$ for $\Delta \tau^+$ and $\Delta \tau^-$, respectively. The error range for $\alpha$ and $\beta$ indicates 90\% confidence interval.}
\begin{ruledtabular}
\begin{tabular}{ccc|ccc|cccccc}
 St & W & Fr~~ &$sign$ & $\alpha$ & $\beta$~~~ & 1 & 2 & 3 & 4 & 5 & 6 \\ \hline 
1 & 0 & $\infty$~~ & + & 1.10$\pm 0.06$ & 1.28$\pm 0.02$~~ & 0.80 & 1.09 & 1.38 & $1.67$ & 1.96 & 2.25 \\
  & & & - & 1.57$\pm 0.08$ & 1.60$\pm 0.02$~~ & 0.68 & 0.91 & 1.14 & 1.38 & 1.61 & 1.85 \\ \hline
4 & 0 & $\infty$~~ & + & 0.98$\pm 0.05$ & 1.20$\pm 0.02$~~ & 0.75 & 1.08 & 1.41 & 1.74 & 2.06 & 2.38 \\
 &  & & - & 1.12$\pm 0.10$ & 1.41$\pm 0.02$~~ & 0.65 & 0.91 & 1.19 & 1.46 & 1.74  &  2.01 \\ \hline
1 & 20 & 0.05~~ & + & 3.06$\pm 0.04$ & 1.80$\pm 0.03$~~ & 0.85 & 1.07 & 1.29 & 1.51 & 1.72 & 1.92 \\
 & & & - & 4.72$\pm 0.18$ & 2.42$\pm 0.06$~~ & 0.77 & 0.92 & 1.07 & 1.21 & 1.35 & 1.49 \\ \hline
4 & 20 & 0.2~~ & + & 2.72$\pm 0.04$ & 1.72$\pm 0.03$~~ & 0.85 & 1.08 & 1.31 & 1.54 & 1.76 & 1.97 \\
 & & & - & 4.24$\pm 0.18$ & 2.38$\pm 0.05$~~ & 0.76 & 0.91 & 1.06 & 1.20 & 1.34 & 1.48 \\
\end{tabular}
\end{ruledtabular}
\end{table*}

In summary, we discovered a vertical strip pattern of particle clustering in our direct numerical simulation of particle-laden turbulence. Quantitative analysis of the averaged distance to the closest particle indicates that the horizontal distance is shorter than the vertical distance, forming a vertical strip pattern only when strong gravity settles particles with St $\ge 1$. A plausible explanation for this phenomenon is provided by the non-Gaussianity or skewness of the flow velocity gradient in the gravitational direction, which is the horizontal convergence, experienced by the particles. The ratio of high-order moments of the persistence time, as shown in Fig. 4, clearly supports our claim that heavy particles settling almost vertically due to strong gravity are more exposed to converging fluid motion, thus forming a vertical strip pattern. Although the effect of gravity on the behavior of particles was investigated before by \cite{Maxey2} in a cellular flow and by \cite{Marchioli, Lavezzo} in channel flows, an identification of the clustering found in our study was difficult due to the unskewed velocity of fluid or inhomogeneity of the flow near the wall, respectively. Finally, it should be mentioned that the parameter range considered in our study is realizable in a real physical problem. According to \cite{Ayala}, the typical range of W (or Fr) for droplets of 10 $\sim$ 60 microns for the dissipation rate $\epsilon = 10 \sim 400~ cm^2/s^3$ in clouds is W = 0.5 $\sim$ 30 (or Fr = 0.01 $\sim 0.2$). This interpretation of the mechanism of clustering due to gravity might be critical for studying how gravity affects droplet collision statistics in cloud turbulence, given that the clustering may enhance droplet coalescence by increasing the collision efficiency\cite{Grabowski, Shaw, Ayala, Onishi}.

This research was supported by a National Research Foundation of Korea (NRF) grant funded by the Korean government (MSIP) (20090093134).


\begin{thebibliography}{00}

\bibitem{Maxey}
M. R. Maxey, J. Fluid Mech. {\bf 174}, 441 (1987).

\bibitem{Squires}
K. D. Squires, J. K. Eaton, Phys. Fluids A {\bf 3}, 1169 (1991).

\bibitem{Reade}
W. C. Reade, L. R. Collins, Phys. Fluids {\bf 12}, 2530 (2000).

\bibitem{Balachandar}
S. Balachandar, J. K. Eaton, Annu. Rev. Fluid Mech. {\bf 42}, 111 (2010).

\bibitem{Monchaux}
R. Monchaux, M. Bourgoin, A. Cartellier,  Int. J. Multiphase Flow {\bf 40}, 1 (2012).

\bibitem{Grabowski}
W. W. Grabowski, L.-P. Wang, Annu. Rev. Fluid Mech. {\bf 45}, 293 (2013).

\bibitem{Jung}
J. Jung, K. Yeo, C. Lee, Phys. Rev. E {\bf 77}, 016307 (2008).

\bibitem{Choi}
Y. Choi, B.-G. Kim, C. Lee, Phys. Rev. E {\bf 80}, 017301 (2009).

\bibitem{Abou1}
A. H. Abdelsamie, C. Lee, Phys. Fluids {\bf 24}, 015106 (2012).

\bibitem{Abou2}
A. H. Abdelsamie, C. Lee, Phys. Fluids {\bf 25}, 033303 (2013).

\bibitem{Lee2004}
C. Lee, K. Yeo, J.-I. Choi, Phys. Rev. Lett. {\bf 92}, 144502 (2004).

\bibitem{ChoiJ}
J.-I. Choi, K. Yeo, C. Lee, Phys. Fluids {\bf 16}, 779 (2004).

\bibitem{Jin}
G. Jin, Y. Wang, J. Zhang, G. He, Ind. Eng. Chem. Res. {\bf 52}, 11294 (2013).

\bibitem{Davila}
J. D\'{a}vila, J. C. R. Hunt, J. Fluid Mech. {\bf 440}, 117 (2001).

\bibitem{Woittiez}
E. J. Woittiez, H. J. J. Jonker, L. M. Portela, J. Atmos. Sci. {\bf 66}, 1926 (2009).

\bibitem{Belin}
F. Belin, J. Maurer, P. Tabeling, H. Willaime, Phys. Fluids {\bf 9}, 3843 (1997).

\bibitem{Batchelor47}
G. K. Batchelor, A. A. Townsend, Proc. R. Soc. Lond. A {\bf 191}, 534 (1947).

\bibitem{Batchelor53}
G. K. Batchelor,  {\it The Theory of Homogeneous Turbulence}, (Cambridge University Press, Cambridge, 1953).

\bibitem{Frisch}
U. Frisch, {\it Turbulence}, 156 (Cambridge University Press, Cambridge, 1995).

\bibitem{Gustavsson}
K. Gustavsson, S. Vajedi, B. Mehlig, arXiv:1401.0513v1 (2014).

\bibitem{Bec}
J. Bec, H. Homann, S. S. Ray, arXiv:1401.1306v1 (2014).

\bibitem{Perlekar}
P. Perlekar, S. S. Ray, D. Mitra, R. Pandit, Phys. Rev. Lett. {\bf 106}, 054501 (2011).

\bibitem{Maxey2}
M. R. Maxey, Phys. Fluids {\bf 30}, 1915 (1987).

\bibitem{Marchioli}
C. Marchioli, M. Picciotto, A. Soldati, Int. J. Multiphase Flow {\bf 33}, 227 (2007).

\bibitem{Lavezzo}
V. Lavezzo, A. Soldati, S. Gerashchenko, Z. Warhaft, L. R. Collins, J. Fluid Mech. {\bf 658}, 229 (2010).

\bibitem{Ayala}
O. Ayala, B. Rosa, L.-P. Wang, W. W. Grabowski, New J. Phys. {\bf 10}, 075015 (2008).

\bibitem{Shaw}
R. A. Shaw, Annu. Rev. Fluid Mech. {\bf 35}, 183 (2003).

\bibitem{Onishi}
R. Onishi, K. Takahashi, S. Komori, Phys. Fluids {\bf 21}, 1251087 (2009).

\end{thebibliography}
\end{document}